\documentclass[page-classic]{villata}

\title{The matter-antimatter interpretation of Kerr spacetime}

\author{M. Villata\thanks{E-mail: \email{villata@oato.inaf.it}}}
\shortauthor{M. Villata}

\institute{INAF, Osservatorio Astrofisico di Torino - Via Osservatorio 20, I-10025 Pino Torinese (TO), Italy}

\abstract{
Repulsive gravity is not very popular in physics. However, one comes across it in at least two main occurrences in general relativity: in the negative-$r$ region of Kerr spacetime, and as the result of the gravitational interaction between matter and antimatter, when the latter is assumed to be CPT-transformed matter. Here we show how these two independent developments of general relativity are perfectly consistent in predicting gravitational repulsion and how the above Kerr negative-$r$ region can be interpreted as the habitat of antimatter. As a consequence, matter particles traveling along vortical geodesics can pass through the throat of a rotating black hole and emerge as antimatter particles (and vice versa). An experimental definitive answer on the gravitational behavior of antimatter is awaited in the next few years.}

\begin{document}

\maketitle

\section{Introduction}

Nearly 100 years ago, on the Russian front, Karl Schwarzschild found the first exact solutions to the Einstein field equations, a few weeks after the publication of the general theory of relativity (and a few months before dying). These solutions describe the geometry of empty spacetime around an uncharged, spherically-symmetric, and non-rotating body \cite{sch16}. The generalization to a charged source was found soon after \cite{rei16,nor18}. However, it took almost fifty years to discover the exact solutions for a rotating, axially-symmetric object, which were found by Kerr in 1963 \cite{ker63}. As usual, the extension to the charged case, i.e.\ the Kerr-Newman solution, followed shortly thereafter \cite{new65}. Further contributions to the investigation of the Kerr metric came in several papers in the following years (e.g.\ \cite{boy65,ker65,car66,boy67,car68,def68}). In particular, efforts were spent to understand the complete topology of Kerr spacetime.

A peculiar aspect of this topology, which was not present in the static Schwarzschild case and in its analytic extensions (e.g.\ \cite{fin58,fro59,kru60,sze60}), is the existence of the negative-$r$ region, i.e.\ the $r$ coordinate is no longer conditioned by the usual restriction $r\ge0$, but is allowed to take also negative values​​, as a Cartesian coordinate. To a physicist this freedom may appear rather embarrassing, also because it is not easy to explain what sense may have this `doubling' of space ​of ​which he had never felt the need before. Here we investigate the properties of this unknown region and seek to give a physical interpretation.

\section{Antimatter and the negative-$r$ side of spacetime}

The metric of a Kerr-Newman spacetime with parameters $M$ (source mass), $a$ (angular momentum per unit mass) and $Q$ (electric charge) in Boyer-Lindquist coordinates $t$, $r$, $\theta$, $\phi$ has the form
\begin{eqnarray}
\label{Eq.1}
{\rm d}s^2=-\left(1-{2Mr\over\rho^2}\right){\rm d}t^2+{\rho^2\over\Delta}\,{\rm d}r^2+\rho^2\,{\rm d}\theta^2+\left(r^2+a^2+{2Mr\over\rho^2}a^2\sin^2\theta\right)\sin^2\theta\,{\rm d}\phi^2\nonumber\\-{4Mr\over\rho^2}a\sin^2\theta\,{\rm d}t\,{\rm d}\phi\,,
\end{eqnarray}
where
\begin{equation}
\label{Eq.2}
\rho^2=r^2+a^2\cos^2\theta\,,
\end{equation}
\begin{equation}
\label{Eq.3}
\Delta=r^2-2Mr+a^2+Q^2\,.
\end{equation}

The corresponding electromagnetic field tensor, expressed in terms of the vector potential ${\textbf{\em A}}$, is
\begin{equation}
\label{Eq.4}
{\textbf{\em F}}={\textbf{d\em A}}={1\over2}(A_{\beta,\alpha}-A_{\alpha,\beta})\,{\bf d}x^\alpha\wedge{\bf d}x^\beta\,,
\end{equation}
with
\begin{equation}
\label{Eq.5}
{\textbf{\em A}}=-{Qr\over\rho^2}({\bf d}t-a\sin^2\theta\,{\bf d}\phi)\,.
\end{equation}

If the source is electrically neutral, i.e.\ $Q=0$ in Eqs.\ (\ref{Eq.3}) and (\ref{Eq.5}), the electromagnetic field vanishes and the metric becomes the uncharged Kerr metric. If also $a=0$, Eq.\ (\ref{Eq.1}) reduces to the static and spherically-symmetric Schwarzschild solution.

One of the main differences between the Schwarzschild and Kerr (or Kerr-Newman) metrics is of topological nature. Indeed, in the former the $r$ coordinate maintains the usual spherical-coordinate condition $r\ge0$, since the metric fails at $r=0$ in an unavoidable singularity. On the contrary, in the Kerr metric, the so-called \textit{ring} singularity is found at $\rho^2=0$, i.e.\ at $r=0$ and $\theta=\pi/2$, so that if a particle (or a photon) approaches the origin with a different value of $\theta$, it is not forbidden (under certain conditions) to cross it and go into the region of \textit{negative} $r$. In fact, Kerr spacetime is usually considered as a manifold formed by taking the topological product ${\bf R}^2\times S^2$ of a 2-plane where $t$ and $r$ are Cartesian coordinates running from $-\infty$ to $+\infty$ and a 2-sphere on which $\theta$ and $\phi$ are ordinary spherical coordinates.

What could be the physical meaning of the negative-$r$ region is not clear. Often it is considered as \textit{another universe}, but no other clue to its existence is currently known. We want here to try to identify this region with something more familiar, possibly already known.

The only needed assumption, which is also the most natural, is that $|r|$ is always a measure of the distance from the origin, in the only physical spacetime we are aware of. In other words, given any value $r_0>0$, $r=r_0$ and $r=-r_0$ identify two 2-spheres that are coincident in space, but evidently with different properties or different `perspectives' of the external world.

Indeed, when $r$ is negative, in Eq.\ (\ref{Eq.5}) it can be replaced by $-|r|$. As a result, both potential and field are reversed, \textit{as if} the electric charge were the opposite one ($-Q$).

Thus, a charged particle in the negative-$r$ `region', e.g.\ an electron, feels an inverted electromagnetic field with respect to the same ordinary particle with positive $r$, as if it were its antiparticle, e.g.\ a positron. In other words, the existence of negative $r$ is equivalent to (or implies) the existence of antimatter. A particle crossing $r=0$ (and avoiding the ring singularity) would suffer nothing else than a charge conjugation. Not intrinsically, but as `seen' from the original region.

What about the gravitational interaction of these antimatter particles (no matter if charged or uncharged, e.g.\ a neutral antiatom or macroscopic antimatter body)? It is well known that in the negative-$r$ region (at least at large $|r|$) gravity is repulsive, as it can be easily seen from Eqs.\ (\ref{Eq.1}) and (\ref{Eq.3}), where negative $r$ values are equivalent to negative $M$ values. This seems to be consistent with the finding that, under the assumption that antimatter is CPT-transformed matter, the gravitational interaction between matter and antimatter is repulsive \cite{vil11}. Indeed, the geodesic equation
\begin{equation}
\label{Eq.6}
{{\rm d}^2x^\alpha\over{\rm d}\tau^2}=-{\Gamma^\alpha}_{\beta\gamma}{{\rm d}x^\beta\over{\rm d}\tau}{{\rm d}x^\gamma\over{\rm d}\tau}\,,
\end{equation}
where the affine connection ${\Gamma^\alpha}_{\beta\gamma}$ represents the (matter-generated) gravitational field and the other three elements refer to the test matter particle, has the property that each of its elements changes sign under a CPT transformation ($q\rightarrow-q$, ${\rm d}x^\alpha\rightarrow-{\rm d}x^\alpha$). If we CPT-transform all the four elements, we obtain an identical equation describing the motion of an antimatter test particle in an antimatter-generated gravitational field, since all the four changes of sign cancel one another. Thus, this CPT symmetry ensures the same self-attractive gravitational behavior for both matter and antimatter. However, if we transform only one of the two components, either the field ${\Gamma^\alpha}_{\beta\gamma}$ or the particle (represented by the remaining three elements), we get a change of sign that converts the original gravitational attraction into repulsion, so that matter and antimatter repel each other.

We have seen that a particle in the negative-$r$ region can be interpreted as its antiparticle in the ordinary $r>0$ region, i.e.\ as transformed by the charge conjugation C. To be fully consistent with the above CPT assumption ensuring repulsive gravity, at first sight it seems that we lack the PT transformation, i.e.\ the fact that in the negative-$r$ region, where antimatter `lives', spacetime is reversed. Actually, the above topology in which $r$ is a Cartesian coordinate implies that there are two symmetry axes $\sin\theta=0$: the north axis ($\theta=0$) and the south axis ($\theta=\pi$), with the latter associated to the negative-$r$ region \cite{haw73,one95}. In addition, the form of the metric (\ref{Eq.1}) shows that the double sign change ${\rm d}t\rightarrow-{\rm d}t$, ${\rm d}\phi\rightarrow-{\rm d}\phi$ is an isometry, which, together with the above equatorial isometry ($\theta\rightarrow\pi-\theta$) of the south axis, can provide the requested spacetime inversion.

What about the electromagnetic properties in this reversed spacetime? As expected, the related PT transformation will invert four-vectors and will leave even-rank tensors unaltered (see \cite{vil11}), so that the field tensor will remain reversed by the former C operation, while the vector potential will invert again, thus returning as it was before the charge conjugation. Indeed, the physically significant quantity is the field, which must eventually be inverted.

For a final check, we can consider the complete equation of motion including the Lorentz-force law
\begin{equation}
\label{Eq.7}
{{\rm d}^2x^\alpha\over{\rm d}\tau^2}=-{\Gamma^\alpha}_{\beta\gamma}{{\rm d}x^\beta\over{\rm d}\tau}{{\rm d}x^\gamma\over{\rm d}\tau}+{q\over m}{F^\alpha}_\beta{{\rm d}x^\beta\over{\rm d}\tau}\,,
\end{equation}
where all elements are CPT-odd, and it is evident that both the gravitational and the Lorentz forces invert when matter and antimatter interact, i.e.\ when one of the two components (either the particle or the fields) is CPT-transformed.

However, whereas spacetime reversal appears to be an essential ingredient for gravitational repulsion between matter and antimatter in the general case of Eqs.\ (\ref{Eq.6}) and (\ref{Eq.7}), it does not seem to be in the highly symmetric case of Kerr-Newman spacetime, where it may be hidden by the above isometries. Alternatively, we can suppose that the somewhat exotic topology involving the negative-$r$ side of spacetime works just as well, as we have seen, in predicting repulsive gravity for antimatter, effectively replacing the CPT transformation.

Actually, both answers are nearly correct. The key to the solution can be found in the Kerr-Newman metric expressed in terms of Kerr-star coordinates:\footnote{Star-Kerr coordinates would work in a similar way. Also, the Kerr metric with $Q=0$ does not imply significant changes in this regard.}
\begin{eqnarray}
\label{Eq.8}
{\rm d}s^2=-\left(1-{2Mr-Q^2\over\rho^2}\right){{\rm d}t^*}^2+\rho^2\,{\rm d}\theta^2+\left(r^2+a^2+{2Mr-Q^2\over\rho^2}a^2\sin^2\theta\right)\sin^2\theta\,{{\rm d}\phi^*}^2\nonumber\\-2{2Mr-Q^2\over\rho^2}a\sin^2\theta\,{\rm d}t^*\,{\rm d}\phi^*+2\,{\rm d}t^*\,{\rm d}r-2a\sin^2\theta\,{\rm d}\phi^*\,{\rm d}r\,,
\end{eqnarray}
where
\begin{equation}
\label{Eq.9}
{\rm d}t^*={\rm d}t+{r^2+a^2\over\Delta}\,{\rm d}r\,,\quad{\rm d}\phi^*={\rm d}\phi+{a\over\Delta}\,{\rm d}r\,.
\end{equation}

The most striking difference with the Boyer-Lindquist form (\ref{Eq.1}) is that the ${\rm d}r^2$ term (which contained the metric singularity for $\Delta=0$) has been replaced by the last two terms in Eq.\ (\ref{Eq.8}).

We want now to demonstrate that the CPT transformation requested in ${\bf R}^4$ to get gravitational repulsion between matter and antimatter [as shown in Eqs.\ (\ref{Eq.6}) and (\ref{Eq.7})] is exactly equivalent to the introduction of negative $r$ of the ${\bf R}^2\times S^2$ topology of Kerr(-Newman) spacetime.

Let us consider the Kerr-star metric (\ref{Eq.8}) as the metric for matter in ${\bf R}^4$ (as in the Schwarzschild case), i.e.\ let us disregard for the moment the existence of the negative-$r$ region and its related antimatter. We know that a CPT transformation in ${\bf R}^4$ represents the existence of antimatter, so that we can extend the matter metric to antimatter by performing on it the transformation and requiring that the form of the metric is unchanged. The PT operation acts as ${\rm d}t^*\rightarrow-{\rm d}t^*$, ${\rm d}\phi^*\rightarrow-{\rm d}\phi^*$, $\theta\rightarrow\pi-\theta$, which correspond to the isometries that left the metric (\ref{Eq.1}) invariant. But in the metric (\ref{Eq.8}) this is no longer true: to get invariance we must also perform the inversion ${\rm d}r\rightarrow-{\rm d}r$, i.e.\ $r\rightarrow-r$. The net result is the extension of the $r$ coordinate to negative values, which gives rise to the complete Kerr(-Newman) metric in the ${\bf R}^2\times S^2$ topology, with the negative-$r$ side pertinent to antimatter. In other words, the exotic topology could have been discovered by means of this PT transformation. The charge conjugation C does not affect the metric, but tells us that we are dealing with antimatter, as we already know from the electromagnetic properties in the negative-$r$ region.

This `topological trick' is also visible in the explicit expression of the electromagnetic field tensor in both coordinate systems. For example, in the Boyer-Lindquist form
\begin{equation}
\label{Eq.10}
{\textbf{\em F}}={Q\over\rho^4}\{(r^2-a^2\cos^2\theta)\,{\bf d}r\wedge({\bf d}t-a\sin^2\theta\,{\bf d}\phi)-2ar\sin\theta\cos\theta\,{\bf d}\theta\wedge[a\,{\bf d}t-(r^2+a^2)\,{\bf d}\phi]\}\,,
\end{equation}
it is evident that the above PT operation (${\rm d}t\rightarrow-{\rm d}t$, ${\rm d}\phi\rightarrow-{\rm d}\phi$, $\theta\rightarrow\pi-\theta$) is equivalent to (and hence can be replaced by) the $r\rightarrow-r$ inversion, which in turn implies the existence and properties of antimatter, as shown at the beginning of this section. The same applies to the vector potential (\ref{Eq.5}).

In summary, the ${\bf R}^2\times S^2$ topology including the $r<0$ side is fully satisfactory in describing both matter and antimatter and their gravitational and electromagnetic interactions, and it is completely consistent with the more general theory of repulsive gravity between matter and CPT-transformed matter.

\section{Examples and conclusions}

To illustrate the physical meaning of our antimatter interpretation of the negative-$r$ side of the Kerr(-Newman) solution of the Einstein(-Maxwell) equations, let us consider axial geodesics in an uncharged Kerr spacetime with $a^2<M^2$, i.e.\ that of a rotating black hole with two horizons at $r=r_\pm=M\pm\sqrt{M^2-a^2}$.

Axial geodesics (i.e.\ those running along the symmetry axis) represent simple examples of different orbit types. When the energy of the particle is greater than the escape energy, they are a particular case of vortical geodesics \cite{def72}, i.e.\ those with Carter constant ${\cal Q}<0$, having the property to pass unimpeded through the throat $r=0$ from the positive-$r$ to the negative-$r$ region. However, not all vortical geodesics have transit orbits from $r=\pm\infty$ to $r=\mp\infty$, but can have flyby orbits with turning point at some $r<0$, whether they come from $r=+\infty$ or $r=-\infty$.

This can easily be shown in the foretold simple case of axial geodesics. On the axis the geodesic equation for $r$ has the form of an energy equation (see e.g.\ \cite{car66}):
\begin{equation}
\label{Eq.11}
\left({{\rm d}r\over{\rm d}\tau}\right)^2+V^2(r)=E^2\,,
\end{equation}
where $E$ is a first integral representing the energy per unit mass and
\begin{equation}
\label{Eq.12}
V^2(r)={\Delta\over r^2+a^2}=1-{2Mr\over r^2+a^2}
\end{equation}
plays the role of a squared potential.

\begin{figure}
\includegraphics[trim = 0 7cm 0 7.5cm, clip, width=\columnwidth]{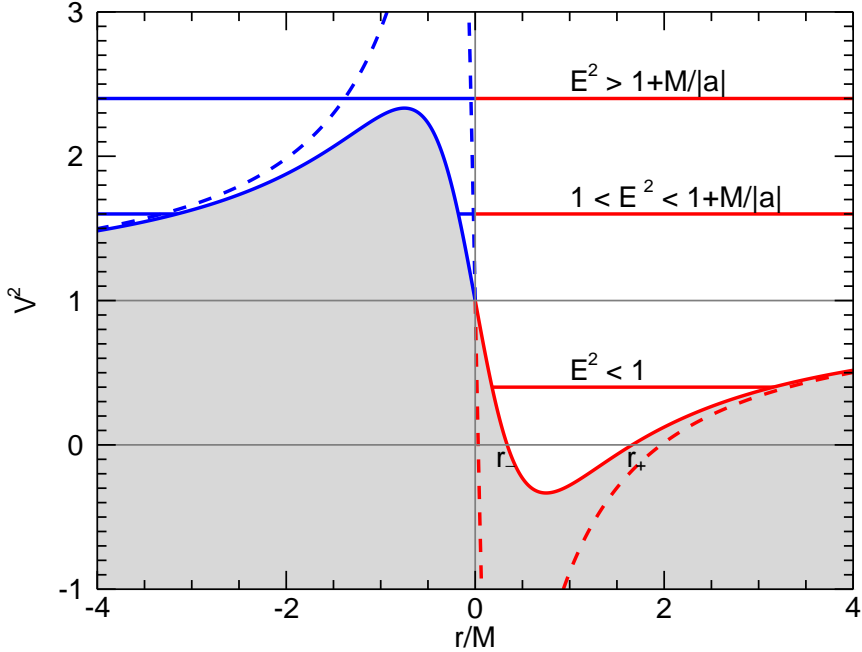}
\caption{The `squared potential' $V^2$ of Eq.\ (\ref{Eq.12}) as a function of $r$ in the interval $-4M\le r\le4M$ with $|a|/M=0.75$ (solid curve). The positions of the two horizons at the intersections of the $V^2=0$ level are indicated as $r_+$ and $r_-$. Horizontal lines represent various $E^2$ levels characterizing different orbit types. For comparison, the trend of $V^2(r)$ with $|a|/M=0.25$ is displayed as a dashed curve.\label{Fig.1}}
\end{figure}

In Fig.\ \ref{Fig.1} we plot $V^2$ with $|a|/M=0.75$ (solid curve) as a function of $r$ in the interval $-4M\le r\le4M$. The positions of the two horizons at the intersections of the $V^2=0$ level (i.e.\ where $\Delta=0$) are indicated as $r_+$ and $r_-$. Horizontal lines represent various $E^2$ levels characterizing different orbit types. For comparison, another example with $|a|/M=0.25$ is plotted as a dashed curve.

When $E^2<1$ we have a bound orbit, with the particle oscillating on the axis in the $r>0$ region through different Kerr patches, since each horizon can be crossed only once. When $E^2$ exceeds 1 we have vortical geodesics, as told above. In this case, if $E^2$ is less than the maximum height of the potential barrier at $r=-|a|$, i.e.\ $1<E^2<1+M/|a|$ ($=7/3$ with $|a|/M=0.75$), particles have flyby orbits with turning point in the negative-$r$ region, at $-|a|<r<0$ for particles coming from $r>0$, closer to the throat as energy decreases towards 1, while for particles in the negative-$r$ region the turning point goes from $-|a|$ to $-\infty$ with decreasing energy, and particles with $E^2<1$ cannot exist there. Also in this case of flyby orbits, particles coming from $r>0$ cannot cross the horizons more than once, so that, after the turn, they are destined to visit another Kerr patch.

In general, particles falling from $r>0$ accelerate under the action of an attractive force down to the potential minimum at $r=|a|$, where they start to slow down due to the effect of an opposite repulsive gravitational field. In the vortical case, the turning point occurs beyond the throat, so that the same repulsive force actually re-attracts towards $r=0$, after which the particle accelerates outwards to cross a new horizon. Particles on flyby orbits at $r<-|a|$ just suffer the repulsive force that prevents them from reaching the throat and that indefinitely accelerates them to their maximum velocity as $r\rightarrow-\infty$.

Finally, particles with $E^2>1+M/|a|$ have transit orbits, thus running unimpeded from $r=\pm\infty$ to $r=\mp\infty$. They achieve their maximum and minimum velocities at $r=|a|$ and $r=-|a|$, respectively, while have the same $|{\rm d}r/{\rm d}\tau|>\sqrt{M/|a|}$ at $r=\pm\infty$.

As shown by the dashed line representing $V^2(r)$ with $|a|/M=0.25$, as $|a|\rightarrow0$, $r_-\rightarrow0$ and $r_+\rightarrow2M$, the potential well becomes infinitely deep and the barrier infinitely high and steep at $r=0$, as in the static Schwarzschild case, where particles of any energy falling from $r>0$ are doomed to hit the central unavoidable singularity. However, there is no reason why the negative-$r$ region should disappear. Although any passage between the two regions is prevented, the negative-$r$ region will continue to exist with its possible population of particles persistently repelled by the `white-hole' nature of the singularity, with the most energetic ones arriving closest to it, but never getting there. Of course, in the static case these axial motions can be referred to any radial direction.

All the above discussion on the axial motion of particles around a rotating black hole is within the standard mathematical frame of Kerr spacetime, where, however, the physical identification of the negative-$r$ region is not clear. We assumed that it does not refer to a different `semi-spacetime' or `another universe', but that it coincides with the known physical spacetime, just presenting a different perspective of it, i.e.\ the perspective of spacetime as seen by antimatter. The mathematical and physical consistency of this scenario has been investigated and discussed in the previous section, showing that particles at $r<0$ behave exactly as expected from the corresponding antimatter particles at $r>0$. Obviously, on their side, antimatter particles can be considered as perfectly normal matter particles, but in a spacetime generated by an antimatter source, i.e.\ with inverted electromagnetic and gravitational fields.

Thus, in our view, particles on the negative-$r$ side of Fig.\ \ref{Fig.1} behave exactly as predicted by the standard Kerr metric, but do it in our spacetime and this is the reason why we can label them as antimatter. Moreover, a particle with transit orbit falling into the black hole after crossing horizons is not lost. Indeed, it will emerge as its antiparticle, without having undergone any intrinsic change of its properties, but having acquired a different, reversed perspective of the geometry of the external world. The energy required for such a transition (in either direction) is minimal in the extreme Kerr black hole $a^2=M^2$, i.e.\ $E^2>2$ (and even lower for naked singularities), while it becomes arbitrarily high as $a\rightarrow0$.

Therefore, in a matter-dominated part of the Universe as those we know, where matter rotating black holes can form and antimatter is originally substantially absent, these objects can work as antimatter factories and the ratio between particles and antiparticles in the environment can change above a given energy which depends on the $a/M$ ratio. This mechanism could play a role in the increasing positron fraction at high energies clearly detected by the Alpha Magnetic Spectrometer on the International Space Station \cite{agu13} and confirming previous results from Fermi-LAT, PAMELA, and other experiments.

This interpretation of Kerr spacetime in terms of matter and antimatter is so simple and natural that may appear very strange that it has never been proposed before (but see \cite{cha97}). Most likely the main reason is that most people have always believed that gravity can only be attractive and the concept of gravitational repulsion between matter and antimatter is rather unpopular and was opposed in the past by some theoretical arguments, which, however, have been subsequently criticized and questioned, and now appear unconvincing (see e.g.\ \cite{vil11,haj11} and references therein).

Now we are confronted with two independent developments of general relativity, the Kerr metric and the theory of CPT gravity, which both predict this gravitational repulsion, in perfect consistency between them.

Within a few years the ongoing experiments at CERN (AEGIS \cite{kel08}, ALPHA \cite{alp13} and GBAR \cite{per12,ind14}) will give a definitive answer on the gravitational behavior of antimatter.

If repulsive gravity between matter and antimatter is confirmed, this discovery will represent a historic breakthrough dense with consequences. As shown in \cite{vil13} (see also \cite{phi14} and references therein), it can explain many of the currently most debated mysteries in physics and cosmology, such as dark energy and dark matter and the cosmic asymmetry between matter and antimatter.


\end{document}